\documentclass{aa}
\usepackage{graphicx,natbib,epsfig}

\bibpunct{(}{)}{;}{a}{}{,} 
\newcommand{\lya}{\hbox{Ly$\alpha$}}

\newcommand{\etal}{\hbox{et al.\ }}
\begin{document}
\title{The near-IR properties and continuum shapes of high redshift quasars from the Sloan Digital Sky Survey}
 \author{Laura Pentericci\inst{1} \and Hans W. Rix\inst{1} \and Francisco Prada\inst{2}
\and Xiaohui Fan\inst{3} \and Michael A. Strauss\inst{4} \and Donald P. Schneider\inst{5}  \and Eva K. Grebel\inst{1} \and Daniel Harbeck\inst{1} \and Jon  Brinkmann\inst{6} \and Vijay K.
Narayanan\inst{4}
}
\institute{ Max-Planck-Institut fur Astronomie, Konigstuhl 17,
             D-69117, Heidelberg, Germany
\and Instituto de Astrofisica de Canarias, E-38205 La Laguna, Tenerife, Spain;
\and Steward Observatory, The University of Arizona
933 N. Cherry Ave, Tucson, AZ 85721-0065  Arizona
\and Princeton University Observatory, Princeton,
 08544 USA
\and Department of Astronomy and Astrophysics, The Pennsylvania State University, University Park, PA 16802
\and Apache Point Observatory P. O. Box 59,
Sunspot, NM 88349-0059 USA
}

\offprints{L.Pentericci, \email{laura@mpia.de}}
  \date{Received date / Accepted date}

\abstract{
We present J-H-K$'$ photometry for a sample of 45 high redshift
quasars found by the Sloan Digital Sky Survey.
The sample was originally selected on the basis of optical
colors and spans a redshift range from 3.6 to 5.03. Our photometry reflects the rest-frame SED longward of Ly$\alpha$ for all redshifts.
The results show that the near-IR colors
 of high redshift quasars are quite uniform.
We have modelled the continuum shape of the quasars (from just
beyond \lya\ to $\sim$ 4000 \AA) with a power law of the form
f$_\nu \propto \nu^\alpha$, and find  $\langle \alpha \rangle
=-0.57$ with a scatter of 0.33. This value is similar to what is
found for lower redshift quasars over the same restframe
wavelength range, and we conclude that there  is hardly any
evolution in the continuum properties of optically selected
quasars up to redshift 5. The spectral indices found by combining
near-IR with optical photometry are in general consistent but
slightly flatter than  what is found for the same quasars using
the optical spectra and photometry alone, showing that the
continuum region used to determine the spectral indices can
somewhat influence the results. 

  }
 \titlerunning{Near-IR properties of high-z QSOs}
 \maketitle
\section{Introduction}
The last few years have seen a dramatic increase in the number
of high redshift optically selected quasars by several surveys
(e.g. Warren \etal\ 1994, Kennefick \etal\ 1995).
Most prominent amongst these surveys
is the Sloan Digital Sky Survey (SDSS -- Fukugita \etal\ 1996, Gunn \etal\ 1998, Hogg \etal\ 2001, York \etal\ 2000, Smith \etal\ 2002, Stoughton \etal\ 2002, Pier \etal\ 2002, Richards \etal\ 2002), which  has
amongst its scientific aims, the construction of the largest
sample of quasars ever, with more than 10$^5$ objects
spanning a large range of redshift and luminosities.
The SDSS has discovered an unprecedented number of new high redshift
quasars,
including more than 200 new quasars at z$\ge$4 (e.g. Fan \etal\ 2001a, 2000, 1999; Zheng \etal\ 2000, Anderson \etal\ 2001, Schneider \etal\ 2001) and the most distant
quasar known to date  at z$=$6.4 (Fan \etal\ 2003, in press).
These high redshift quasars have been efficiently
selected by their distinctive position in
 color-color diagrams, with characteristic colors due
to the main features of the quasar
 spectra, viz., the power law continuum, the
strong Ly$\alpha$ emission line,
the Ly$\alpha$ forest absorption and Lyman limit, all of which
move quasars away from the stellar color locus.
\\
In this paper we aim
to determine the continuum properties of high redshift quasars at optical restframe wavelengths.
To do this we have obtained
J-H-K$'$ photometry of a large color-selected sample
 of high redshift quasars found by SDSS.
A good knowledge of the quasar continuum shape at near-UV/optical restframe
wavelengths is important for several reasons.
Over the near-UV/optical wavelength range
the shape of the continuum is usually approximated  with a single
power law of the form $f(\nu)\propto \nu^{\alpha}$. The mean value of $\alpha$,
the dispersion of this value, and even the validity of the
power law parameterization as well as the evolution
of such parameters with redshift are still under debate (e.g. Vanden Berk \etal\ 2001 and references therein).
The continuum slopes of quasars are blue,
with a mean canonical spectral index of $\alpha = -$0.7 (e.g. Richstone and Schmidt 1980),
but several results of the past few years point to flatter continua
(e.g. Francis 1996,
Natali \etal\ 1998, Vanden Berk \etal\ 2001),
and indicate a different slope
at different restframe wavelength ranges.
On the other hand the most recent results of Fan \etal\ (2001a) and Schneider \etal\ (2001), who find
steeper average indices (respectively $\alpha =-$0.8 and $-$0.9)
for very high redshift quasars, seem to point to an evolution of the continuum properties of quasars with redshift.
However, as discussed in Schneider \etal\ (2001), for high redshift quasars,
determination of the continuum slope from optical data alone ($\lambda_{obs} < 1 \mu$) must rely on a small restframe wavelength region with $\lambda_{rest} < 1800-1900$ \AA.
For example in objects
at z$\sim$4, the restframe V-band emission is shifted to K-band. Therefore near-IR
data are essential to  unambiguously determine the continuum properties of high redshift quasars in a way comparable to low-redshift objects, so as to get an unbiased measure of evolution.
\\
A  knowledge of the optical continuum shape and its possible redshift evolution,
is not only important to understand the quasars itself but
also for selecting objects at  even higher
redshifts. Optical colors alone begin to be less efficient for selection purposes at z$\sim$ 5, since the quasars
evolutionary track crosses the locus of very
low-mass, late-type stars.
Already, the highest redshift quasars discovered to date have been selected by adding J-band photometry
 to the SDSS colors,
to distinguish quasar candidates from stars
(see in particular  Zheng \etal\ 2000 and Fan \etal\ 2001a).
\\
In addition, determining
the flux decrement in the \lya\ forest needs an
estimate of the continuum at $\lambda < \lambda_{Ly \alpha}$,
which is usually an extrapolation from $\lambda > 1 \mu$m.
If this is done by using the classical $\alpha=-0.7$ index
it can lead to uncertainties of 5-10\% in the
computation of the continuum decrement.
Also for high redshift quasars the observed optical fluxes must be extrapolated to obtain the optical luminosity M$_B$, e.g. to determine the luminosity function. Any change of $\alpha$ with redshift could substantially influence the inferred luminosity. Only with near-IR photometry can
M$_B$ be determined directly. Lastly, for the SDSS quasar search,
a knowledge of the continuum slope is essential for
modeling the SDSS quasar selection function which is then used to derive the luminosity function (Fan \etal\ 2001b).
\section{Sample selection}
High-redshift (z$>$ 3.6) quasar candidates were selected using color cuts
that separate them from the stellar locus
(e.g. Fan \etal\ 2000, 2001a).
For the color-selected statistical sample presented in \S 4, the
color criteria were applied  {\em after} correcting for Galactic
extinction using the reddening map of Schlegel, Finkbeiner \& Davis (1998).
The criteria are as follows:

1. $gri$ candidates, selected principally from the $g^*-r^*, r^*-i^*$ diagram:

\begin{equation}
 \begin{array}{l}
        (a)\ i^* < 20.05 \\
        (b)\ u^* - g^* > 2.00 \mbox{ or } u^* > 21.00 \\
        (c)\ g^* - r^* > 1.00 \\
        (d)\ r^* - i^* < 0.42 (g^* - r^*) - 0.31 \mbox{ or } g^*- r^* >2.30 \\
        (e)\ i^* - z^* < 0.25
        \end{array}
\end{equation}

2. $riz$ candidates, selected principally from the $r^*-i^*, i^*-z^*$ diagram:

\begin{equation}
\begin{array}{l}
        (a)\ i^* < 20.20  \\
        (b)\ u^* > 22.00 \\
        (c)\ g^* > 22.60 \\
        (d)\ r^* - i^* > 1.00 \\
        (e)\ i^* - z^* < 0.47 (r^* - i^*) - 0.48
        \end{array}
\end{equation}

The intersections of those color cuts with the $g^* - r^*, r^*-i^*$ and
$r^*-i^*,i^*-z^*$ diagrams are illustrated e.g. in Fan \etal\ (2000) and (2001a).

Within these color magnitude boundaries, two subsamples were selected for observations in Spring and Fall.
The first is the color selected sample
in the Fall Equatorial Stripe, presented in Fan \etal\ (2001a),
which comprises 39 objects.
The second is  a complete  sample in the Spring Equatorial Stripe, consisting
of 55 objects whose redshifts have been reported in
different papers (Schneider \etal\ 2001, Fan \etal\ 2000, Fan \etal\ 2003
in preparation , Anderson \etal\ 2001).
Since the two samples have been selected with the same optical color criteria
they can be merged together and form a large color selected sample, spanning a redshift range from 3.60 to 5.03.
\\
The completeness of the Fall sample has been extensively discussed in Fan \etal\  (2001a)
and is around 80 \%, depending slightly on redshift; the completeness  for the Spring sample
should be very similar. As we detail below, we have observed a random subsample of 45 of the 94 quasars. Therefore our sample  can be viewed as a statistical sample with the above
color cuts.
\begin{table*}
\begin{center}
\caption{\rm Observations log}
\begin{tabular}{cccccccc } \hline \hline
Run  & date     &seeing  & nights/phot &  t$_J$          &  t$_H$        &  t$_K$ & no of obj    \\
(1)  & (2)      &(3)     & (4)         &(5)&(6)&(7)&(8) \\
\hline
1    & Nov 1999 & 1       &   1/1   & 10m (10sx3) & 10m (10sx3)& 10m (6sx5)  &  8   \\
2    & Aug 2000 & 0.9-1.1 &   7/3   & 12m (10sx6) & 12m (10sx6)& 12m (6sx5)  & 19   \\
3    & Feb 2001 & 1$\leq$ &   4/2   & 12m (15sx8) & 12m (15sx8)& 12m (5sx12) & 17   \\
4    & Mar 2001 &         &   6/0   & 12m (10sx6) & 12m (10sx6)& 12m (10sx6) &  0   \\
5    & Jun 2001 & 1.4-1.5 &   2/1   & 12m (10sx6) & 12m (10sx6)& 12m (5sx12) &  6   \\
6    & Nov 2001 & 1-1.5   &   3/2   & 12m (10sx6) & 12m (10sx6)& 12m (5sx12) & 12   \\
7    & Feb 2002 & 1.3-1.5 &   1/1   & 12m (10sx6) & 12m (10sx6)& 12m (5sx12) &  3   \\
\hline
\end{tabular}
\end{center}
(1) Run identification (2) Observation date (3)  average seeing (FWHM) during the night (4) Total number of nights/ total number of photometric-good nights (5) Total integration time in J-band with sub-integration time and number of frames at each position (6) The same for H-band (7) The same for K-band (9) Total number of objects observed (including repetition and objects that are not in the  present sample)
\end{table*}

\section{Observations and data reduction}
Observations were carried out in service mode with the MAGIC near-IR camera on the 2.2m
telescope located at Calar Alto (Spain). MAGIC
is equipped with a Rockwell 256x256 pixel NICMOS3 detector array.
In the high resolution mode it provides a pixel scale of 0.64$''$ pixel$^{-1}$ and a
total field of view of 164$'' \times$164$''$. The observations log is shown in Table 1.
In this paper we only present data from nights with photometric
or nearly photometric conditions (clear nights, with very
stable conditions).
A fraction of the non-photometric time
was used to obtain J-band snapshots of
candidate high redshift quasars from the SDSS, selected as  i-band dropouts,
for which subsequent spectroscopy was carried out at other telescopes
(e.g. Fan \etal\ 2001b).

Each target was observed through the J, H, and K$'$ filters.
The K' band (Wainscoat \& Cowie 1992)
was preferred to the K band to reduce the effects of thermal emission from the sky
and the telescope.
Each object was observed for a total on-target exposure time between 10 and 12 minutes in each filter.
For a satisfactory background subtraction, the images were
dithered on a 6  position pattern (4 positions for the first run),
with offsets of around 30$''$ between each frame.
We took 6, 12 or 18 frames per object depending on the
night and on the brightness of the object. At each position the total integration time was split into short
sub-frames to avoid saturation (the frames were directly integrated at the telescope).
In Table 1 we report all these observational parameters, namely the total integration time, the sub-integration times and the number of frames at each position.

During each night, 4 or more different standard stars
(UKIRT faint stars from Elias \etal\ 1982, complemented by
those observed in the same field by Hunt \etal\ 1998)
spanning a range of colors
 were observed in each of the filters
at regular intervals (at least 3 times per night).
In most cases the standards span
the same range of airmasses as the targets.
From the fall sample a total of 29 objects (out of 39) were observed
under photometric or nearly photometric conditions.
From the spring sample a total of 16 (out of 55) objects were observed.
The selection of the objects was
only determined by chance and observability so it is unbiased with respect
to quasar properties.
Therefore a total of 45 objects form a
statistically well defined sample with complete near-IR photometry.
\\
Note that the J, H and K$'$ data for a given object were taken on a single night, while
the near-IR data were obtained one to two years after
the optical fluxes were measured from SDSS, so we cannot exclude the possibility that variability
could influence some of the results,
such as the determination of the spectral indices of the continuum slopes.
For example, a variability of 10-20\% in the optical
flux could change the measured slope by $\sim 0.2$.

Data reduction was performed with IRAF\footnote{IRAF is distributed by the National Optical Astronomy Observatory, which is operated by the AURA, Inc under cooperative agreement with the NSF.}.
The  sky-background was subtracted using a composite
sky frame made for each image from the previous and following images.
Typically five neighboring frames were used to create a sky image,
although in a few cases when the sky was
changing more rapidly we used only three frames.
Dome flats were obtained by
taking exposures with dome lamps on and off and subtracting one from the other; these were applied after sky subtraction (an alternative flat field was also created by averaging a large number of sky-frames,
masking out the brightest stars).
The background subtracted, flatfielded images were
shifted to a common reference, with shifts derived
from centering the position of as many point sources
as were visible in the frames.
In most cases at least four or more such reference stars were present,
but in a few objects only one reference could be used.
The registered images were then co-added using the IRAF/avsigclip
rejection algorithm.
With the large number of frames for each target (typically 12),
it was easy to perform cosmic ray rejection.

Photometric calibration was done using at
least five standard stars to derive the zero point of the three
bands. During each good night the scatter in the photometric zero point
was less than 0.05 (slightly higher for J band), without color correction so
all the standard stars in a given
band were simply averaged to give the final calibration.
The photometry is accurate to 0.1 magnitudes for the bright
part of the sample.
Magnitudes were derived inside circular apertures with radius of 4$''$.
The K$'$ filter, as opposed to the standard K filter, does not
have many published values for standard star photometry,
but for some stars
interpolated K$'$ magnitudes were available
from  the web page developed by Dave Thompson\footnote{http://www.astro.caltech.edu/mirror/keck/realpublic/inst/ \\ nirc/UKIRTstds.html}.
For those stars which did not have K$'$ magnitudes
available we simply used the K band magnitude
from Hunt \etal\ (1998).

 \begin{figure}
\psfig{figure=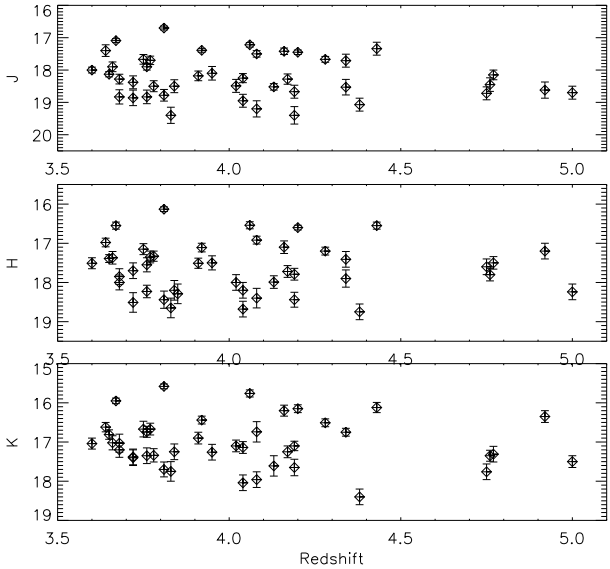,width=9.5cm}
 \caption{The J,H and K magnitude of the quasars plotted as a function of
redshift, with photometric uncertainties indicated by the errorbars.}
\end{figure}
  \begin{figure}
\psfig{figure=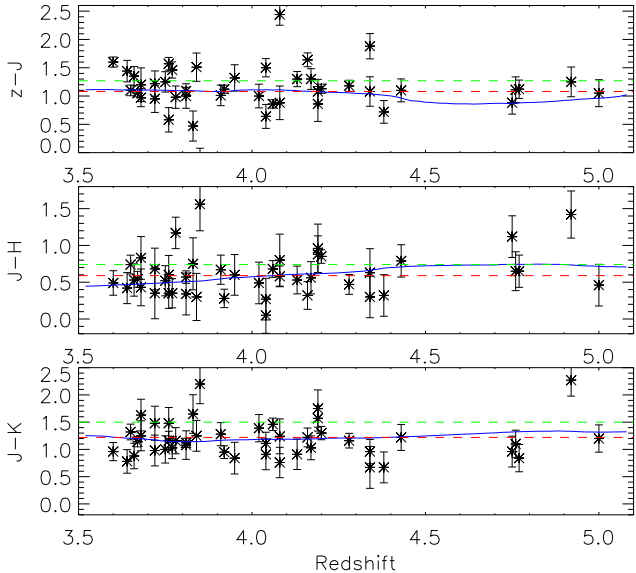,width=9.1cm}
 \caption{The near-IR colors of the sample quasars plotted as a function of redshift.
The dashed lines are the color expected from an object with a
 power law continuum of slope  $\alpha=-0.5$ (below) and $\alpha=-1.0$ (above), while the solid lines are the colors
obtained by redshifting the composite SDSS quasar spectrum (Vanden Berk \etal\  2001).}
\end{figure}

\section{Results and discussion}
\subsection{Colors}
The resulting J, H and K$'$ magnitudes with their corresponding  errors
are presented in Table 2. The photometric errors are dominated
by the sky noise.
In Figure 1 we show the distribution of J, H and K$'$ magnitudes with redshift for all quasars in the sample.
In Figure 2 we show their colors (z$-$J, J$-$H and  J$-$K$'$)
again plotted as a function of redshift.
Note that the z band values are from SDSS photometry, based on the
AB magnitude system, while the IR-colors are on the Johnson (Vega based) system.
In Figure 3 we present the shape of the spectral energy distributions (SED) for all objects, combining the SDSS five-band photometry (from Fan \etal\ 2001a; 2000 and Richards \etal\ 2001) with our near-IR colors.
The observed flux points have been shifted to the restframe wavelength and normalized to have the same i-band magnitude (the i-band flux being of higher S/N than the z band).
For these plots we have consistently
converted the near-IR photometry into AB magnitudes using the following zero points: J$_{AB}=J+0.89$,
H$_{AB}=H+1.38$ and  K$'_{AB}=K'+1.84$.
These offsets were derived assuming that the flux density from Vega is constant within any filter and
equal to 1620 , 1020 and 688 Jy:
the first two values are reported on the UKIRT homepage\footnote{http://www.jach.hawaii.edu/JACpublic/UKIRT/astronomy/ \\ conver.html},
while for the K$'$ filter we used an extrapolation from a measurement available in the MAGIC homepage\footnote{http://www.caha.es/CAHA/Instruments/IRCAM/MAGIC/}.
\\
In Figure 2 we have also plotted as dashed  lines
the colors expected for a quasar if the continuum were
a perfect power law with a spectral index $\alpha=-0.5$ and $\alpha=-1.0$
respectively. From the continuum only,
the colors expected would be
z$-$J$=$1.08, J$-$H$=$0.69 and J$-$K$=$1.22 for $\alpha =-0.5$.
The solid lines in each panel represent the colors
produced by shifting the composite SDSS quasar spectrum
by Vanden Berk \etal\ (2001) to the different redshifts.
This spectrum is best represented by a slightly
flatter power law continuum ($\alpha=-0.46$) and of course contains the contribution
of all emission lines and features.
The two lines deviate substantially from each other only in the z$-$J color
for objects at redshift $>$4.5, due to the contribution of CIV entering
the z-band, and in the J$-$H color for redshift less than $\sim $ 4,
due to the
contribution of the MgII and the FeII complex in J band (Richards \etal\ 2001, Barkhouseand Hall 2001).
\\
From the plots it is clear that the individual quasars show scatter in their colors
 around the predicted mean color
lines. The scatter is far in excess of the measurement errors.
Such scatter is produced by several factors, including
the intrinsic differences of the
spectral properties of quasars, which can have bluer or redder continua and
the different relative strength of the emission lines in different objects. Note that in Figure 2, the object at z$=$4.92 with colors deviating substantially from the expected ones, is a BAL quasar SDSSJ 160501.21$-$011220.6 (Fan \etal\ 2000), which shows quite unusual optical colors (see discussion in Hall \etal\ 2002). Apart from this object and SDSS J103432.72$-$002702.6, which is a mini BAL,
there are no other BAL quasars in the sample.
\\
There are only few near-IR measurements available in the literature for quasars at a similar redshift (e.g. Rodriguez Espinosa \etal\ 1988, Bechtold \etal\ 1994, Francis 1996, Zheng \etal\ 2000), all indicating  colors similar to what we find.
\\
Note that  Zheng \etal\ (2000) used z$-$J$<$ 1.5 and J$-$K $<$1.8 as additional constraints to select
candidate high redshift
quasars, whereas Fan \etal\ (2001b) used only  z$-$J$<$ 1.5.
We see that actually all except 3 of the objects satisfy
the first constraint, and only one does not satisfy the second criterion.
Indeed all L and M dwarfs stars found by combined SDSS plus 2MASS photometry
have z$-$K $>$ 2 (Finlator \etal\ 2000, Leggett \etal\ 2002) so their colors are quite different from those of quasars. These results
 confirm that the addition of one single near-IR band information to the SDSS photometry
can greatly improve the efficiency of finding  high redshift quasars.

\begin{figure}
\psfig{figure=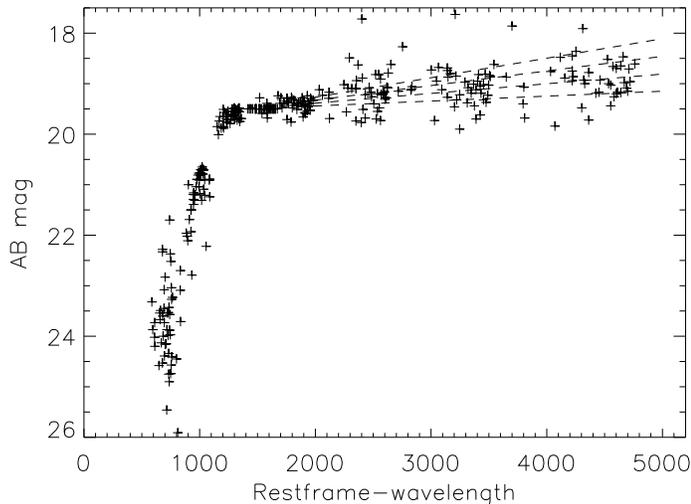,width=10.cm}
 \caption{The spectral energy distribution of all quasars:
all near-IR magnitudes have been transformed into the AB system (see text for details).
The observed values have been shifted to the restframe of each object and normalized to have
 the same i-band magnitude. The lines indicate power laws with spectral indices from -0.25 to -1.0 (from the lower to the upper line). The object with colors much brighter than the other is 2256$+$0047, one of those whose continuum is not well represented by a power-law. It could be also a variable quasar.}
\end{figure}

\begin{table*}
\begin{center}
\caption{\rm Near-IR multiband photometry of high redshift SDSS quasars}
\begin{tabular}{cccccccc } \hline \hline
SDSS name                 & J             & H             & K              & z     & Ref & run & $\alpha$ \\
                          &               &               &                &       &     \\
\hline
SDSS J001950.06$-$004040.9 &17.71$\pm$0.2  &17.41$\pm$0.15 & 16.75$\pm$0.10 & 4.34 & F01 &2 & -0.79\\
SDSS J003525.29$+$004002.8 &--             &17.70$\pm$0.20 & 17.72$\pm$0.15 & 4.75 & F99 &1 & -0.34\\
SDSS J005922.65$+$000301.4 &17.42$\pm$0.10 &17.10$\pm$0.16 & 16.20$\pm$0.14 & 4.16 & F01 &2 & -0.94\\
SDSS J010619.25$+$004823.4 &17.34$\pm$0.08 &16.55$\pm$0.09 & 16.12$\pm$0.13 & 4.43 & F99 &2 & -0.75\\
SDSS J012019.99$+$000735.5 &19.20$\pm$0.35 &18.30$\pm$0.30 & 17.96$\pm$0.30 & 4.08 & F01 &2 & -0.02\\
SDSS J012403.78$+$004432.7 &16.70$\pm$0.06 &16.13$\pm$0.06 & 15.58$\pm$0.06 & 3.81 & F99 &7 & -0.48\\
SDSS J012650.77$+$011611.8 &17.90$\pm$0.15 &17.37$\pm$0.16 & 17.02$\pm$0.15 & 3.66 & F99 &6 & -0.47\\
SDSS J012700.69$-$004559.1 &17.22$\pm$0.07 &16.54$\pm$0.09 & 15.76$\pm$0.09 & 4.06 & F01 &2 & -0.45\\
SDSS J013108.19$+$005248.2 &19.40$\pm$0.27 &18.44$\pm$0.19 & 17.65$\pm$0.20 & 4.19 & F01 &2 & -0.20\\
SDSS J015048.83$+$004126.2 &17.09$\pm$0.06 &16.55$\pm$0.09 & 15.95$\pm$0.07 & 3.67 & F99 &6 & -0.39\\
SDSS J015339.61$-$001104.9 &17.45$\pm$0.07 &16.60$\pm$0.06 & 16.15$\pm$0.10 & 4.20 & F99 &2 & -0.68\\
SDSS J020731.68$+$010348.9 &18.90$\pm$0.20 &18.34$\pm$0.30 & 17.70$\pm$0.30 & 3.85 & F01 &6 & -0.40\\
SDSS J021043.17$-$001818.4 &18.15$\pm$0.13 &17.43$\pm$0.13 & 17.30$\pm$0.20 & 4.77 & F01 &6 & -0.26\\
SDSS J023231.40$-$000010.7 &18.80$\pm$0.11 &18.60$\pm$0.22 & 17.70$\pm$0.19 & 3.81 & F99 &6 & -0.03\\
SDSS J025019.78$+$004650.3 &18.45$\pm$0.21 &17.80$\pm$0.16 & 17.35$\pm$0.14 & 4.76 & F01 &6 & -0.40\\
SDSS J025112.44$-$005208.2 &18.50$\pm$0.17 &17.33$\pm$0.13 & 17.34$\pm$0.17 & 3.78 & F99 &6 & -0.20\\
SDSS J030025.23$+$003224.2 &18.67$\pm$0.20 &17.79$\pm$0.15 & 17.10$\pm$0.12 & 4.19 & F01 &6 & -1.04\\
SDSS J031036.85$+$005521.7 &17.70$\pm$0.13 &17.35$\pm$0.15 & 16.67$\pm$0.15 & 3.77 & F01 &7 & -0.80\\
SDSS J035214.33$-$001941.1 &18.28$\pm$0.12 &17.72$\pm$0.11 & 17.25$\pm$0.10 & 4.17 & F01 &3 & -0.73\\
SDSS J033829.31$+$002156.3 &18.70$\pm$0.20 &18.24$\pm$0.30 & 17.50$\pm$0.30 & 5.00 & F99 &1 & -0.26\\
SDSS J085430.18$+$004213.6 &18.55$\pm$0.10 &17.95$\pm$0.10 & 17.70$\pm$0.20 & 4.13 & A01 &3 & -0.85\\
SDSS J091637.55$+$004734.1 &18.50$\pm$0.16 &17.60$\pm$0.16 & 17.10$\pm$0.10 & 3.72 & F03 &6 & -0.38\\
SDSS J095755.64$-$002027.5 &17.90$\pm$0.10 &17.55$\pm$0.11 & 16.74$\pm$0.10 & 3.76 & F03 &6 & -0.67\\
SDSS J100151.59$-$001626.9 &18.13$\pm$0.08 &17.39$\pm$0.10 & 16.81$\pm$0.12 & 3.65 & F03 &3 & -0.67\\
SDSS J101832.46$+$001436.4 &19.50$\pm$0.30 &18.50$\pm$0.20 & 17.77$\pm$0.21 & 3.83 & S01 &3 & -0.50\\
SDSS J103432.72$-$002702.6 &18.80$\pm$0.20 &18.43$\pm$0.30 & 18.30$\pm$0.30 & 4.38 & F03 &3 &  0.01\\
SDSS J105602.37$+$003222.0 &18.45$\pm$0.15 &18.00$\pm$0.30 & 17.08$\pm$0.18 & 4.00 & F03 &1 & -0.61\\
SDSS J111224.18$+$004630.4 &18.94$\pm$0.20 &18.60$\pm$0.25 & 17.90$\pm$0.25 & 4.04 & A01 &3 &  0.09\\
SDSS J111246.29$+$004957.5 &17.39$\pm$0.06 &17.11$\pm$0.11 & 16.44$\pm$0.10 & 3.92 & F00 &3 & -0.30 \\
SDSS J134723.09$+$002158.9 &17.67$\pm$0.09 &17.20$\pm$0.10 & 16.51$\pm$0.10 & 4.28 & A01 &3 & -0.51\\
SDSS J135134.46$-$003652.2 &18.25$\pm$0.10 &18.25$\pm$0.20 & 17.14$\pm$0.15 & 4.04 & A01 &3 & -0.76\\
SDSS J135828.75$+$005811.5 &18.18$\pm$0.10 &17.51$\pm$0.10 & 16.90$\pm$0.15 & 3.91 & F03 &5 & -0.51\\
SDSS J140754.54$+$001312.1 &18.86$\pm$0.20 &18.50$\pm$0.30 & 17.30$\pm$0.20 & 3.72 & F03 &3 & -0.49\\
SDSS J142329.99$+$004138.4 &18.83$\pm$0.17 &18.20$\pm$0.15 & 17.40$\pm$0.20 & 3.76 & F00 &3 & -0.08\\
SDSS J142647.81$+$002740.1 &18.28$\pm$0.10 &17.80$\pm$0.15 & 17.03$\pm$0.15 & 3.68 & F00 &5 & -0.41\\
SDSS J160501.21$-$011220.6 &18.30$\pm$0.26 &17.20$\pm$0.10 & 16.40$\pm$0.10 & 4.92 & F00 &2 & -1.72\\
SDSS J162500.04$-$000128.5 &18.00$\pm$0.09 &17.51$\pm$0.14 & 17.04$\pm$0.14 & 3.60 & F03 &5 & -1.03\\
SDSS J225624.35$+$004720.2 &17.48$\pm$0.10 &16.90$\pm$0.10 & 16.74$\pm$0.15 & 4.08 & F01 &2 & -1.77\\
SDSS J225759.67$+$001645.7 &17.67$\pm$0.15 &17.15$\pm$0.14 & 16.60$\pm$0.15 & 3.75 & F99 &2 & -0.34\\
SDSS J230323.77$+$001615.2 &18.83$\pm$0.20 &18.25$\pm$0.15 & 17.20$\pm$0.18 & 3.68 & F01 &5 & -0.80\\
SDSS J230639.65$+$010855.2 &17.40$\pm$0.18 &16.98$\pm$0.11 & 16.62$\pm$0.12 & 3.64 & F01 &6 & -0.67\\
SDSS J230952.29$-$003138.9 &18.10$\pm$0.21 &17.50$\pm$0.14 & 17.26$\pm$0.20 & 3.95 & F99 &2 & -0.47\\
SDSS J232208.22$-$005235.2 &18.60$\pm$0.20 &18.14$\pm$0.20 & 16.80$\pm$0.10 & 3.84 & F01 &6 & -0.83\\
SDSS J235718.35$+$004350.4 &18.53$\pm$0.20 &17.97$\pm$0.18 & 17.86$\pm$0.30 & 4.34 & F99 &2 & -0.86\\

                                   \end{tabular}
\end{center}
References for redshift: F99 Fan \etal\ 1999; F01 Fan \etal\ 2001a; F03 Fan \etal\ 2003 in preparation; A01 Anderson     \etal\ 2001; S01 Schneider  \etal\ 2001.
\end{table*}

\subsection{Continuum slopes}
The continuum slopes were computed using all
SDSS bands that fall fully redward of the \lya\ line
(i.e. excluding the band that contains \lya) in combination with
the three near-IR bands.
As already noted in the previous section, contamination
by line emission can be important in some bands.
The most important lines that are present in the quasar
spectra at the restframe
wavelengths sampled by our observations
are the CIV  ($\lambda_{rest} =1549$ \AA) and CIII]  ($\lambda_{rest} =1909$ \AA) lines
for the optical photometry (i and z-band); the
MgII line ($\lambda_{rest} =2798$\AA) which falls in the J-band
for objects with  z$<$ 4.0 and in the
 H-band for objects with z$>$ 4.36;
finally H$\beta$ ($\lambda_{rest} =4861$ \AA) which
contaminates the K$'$ band for objects with z$<$ 3.73.
Many other lines also fall in
these regions but their contributions to broad band fluxes are mostly negligible.
A special case is the very broad feature that appears between 2200-4000 \AA,
 known as the 3000 \AA\ bump, which consists of blends of FeII line emission and
Balmer continuum emission (e.g. Wills \etal\ 1985, Richards \etal\ 2002).
No attempt was made to correct for these features since their
contribution is hard to disentangle.
However, we have seen
from the color distribution in the previous section
that this contribution cannot be very large, or at least it cancels
out, otherwise we would observe larger departures from the values expected from a simple power
law continuum. Furthermore we are using a quite large wavelength baseline (from 7000 to 22000 \AA)
which means that these errors would not have a large effect on the derived continuum slope.
For example an error of 25 per cent in the flux of one of the bands would introduce an error of $\le$0.1 in
$\alpha$ (in the sense that if the
25\% was due to the flux contributed by the FeII bump, the
true $\alpha$ would be flatter than what we measure).
\\
For a large fraction of the  objects in the sample
the strength of the CIV line is known, having been measured from the
low resolution discovery spectra (Fan \etal\ 1999; Fan \etal\ 2001a; Fan \etal\ 2003 in preparation; Anderson \etal\ 2001; Schneider  \etal\ 2001).
The strength of the other lines was not known so we used the average
equivalent width derived from the combined spectra of
SDSS quasars by Vanden Berk \etal\ (2001).
In particular we used EW=23.78 \AA\ for CIV (when the real
value was not available),
EW=21.19 \AA\ for CIII], EW= 34.95 \AA\ for MgII and EW=46.21 \AA\ for H$\beta$
(these are all restframe values).
Note that in most cases the
 contribution of line emission flux to the broad band flux is less than  0.1 magnitudes,
so it is comparable to or less than the photometric error.

We subtracted the line contribution from the relevant broad band values
and then converted magnitudes into flux.
The corrected continuum fluxes were then modeled with a
power law of the form $f_{\nu} \propto \nu^{\alpha}$ and the best fitting
spectral index $\alpha$ was derived by least squares minimization.
The fit by a power law was considered acceptable when the probability that a
value of chi-square as poor as the value found should occur by chance
was larger than 0.05.
For 5 objects in the sample (0019$-$0040, 0035$+$0040, 1625$-$0001, 2256$+$0047 and 2306$+$0108)
this requirement was not satisfied, so we report the
values we have obtained in Table 2, but we do not include them in the discussion.

\begin{figure*}
\psfig{figure=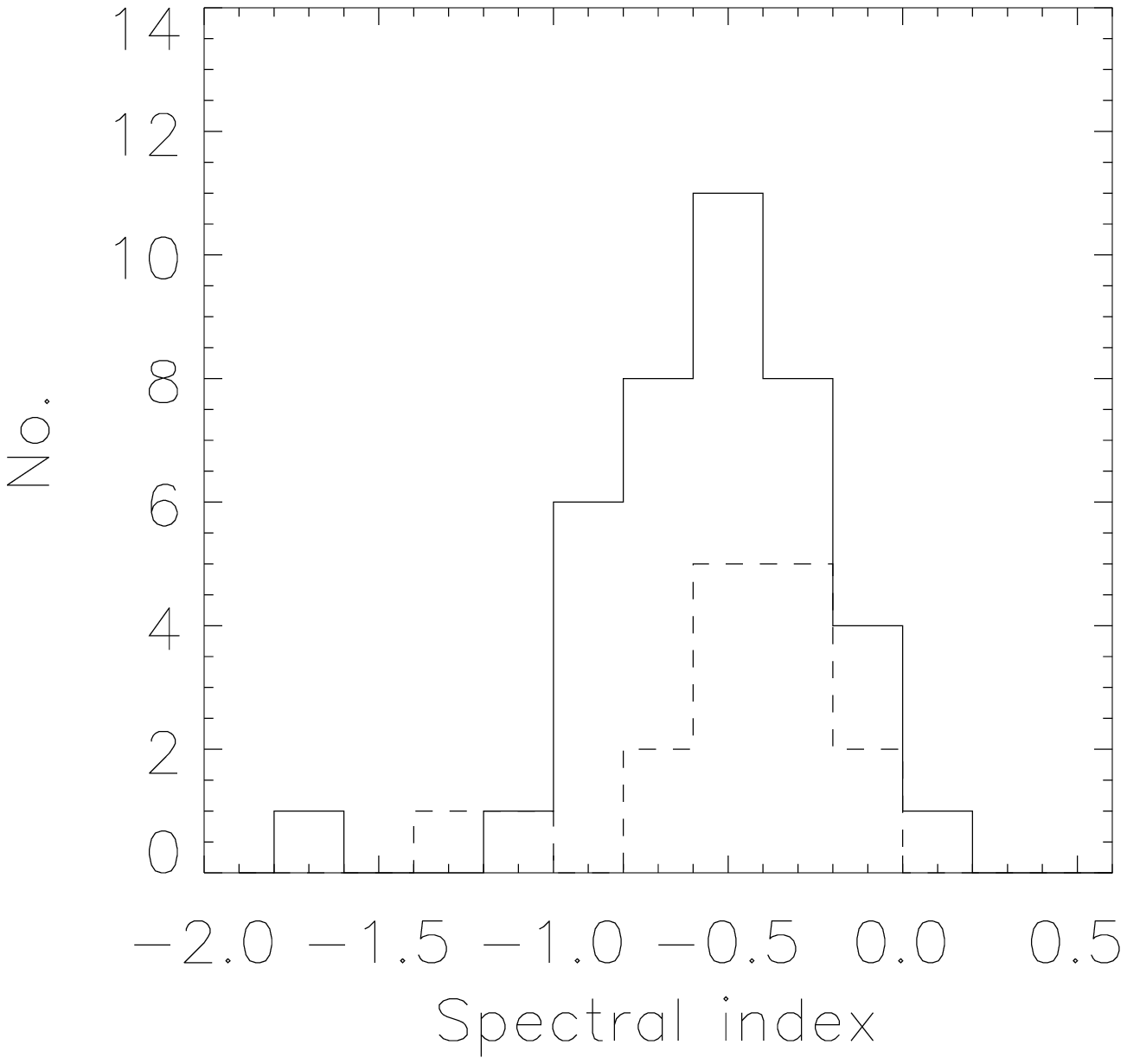,width=8.5cm}
\psfig{figure=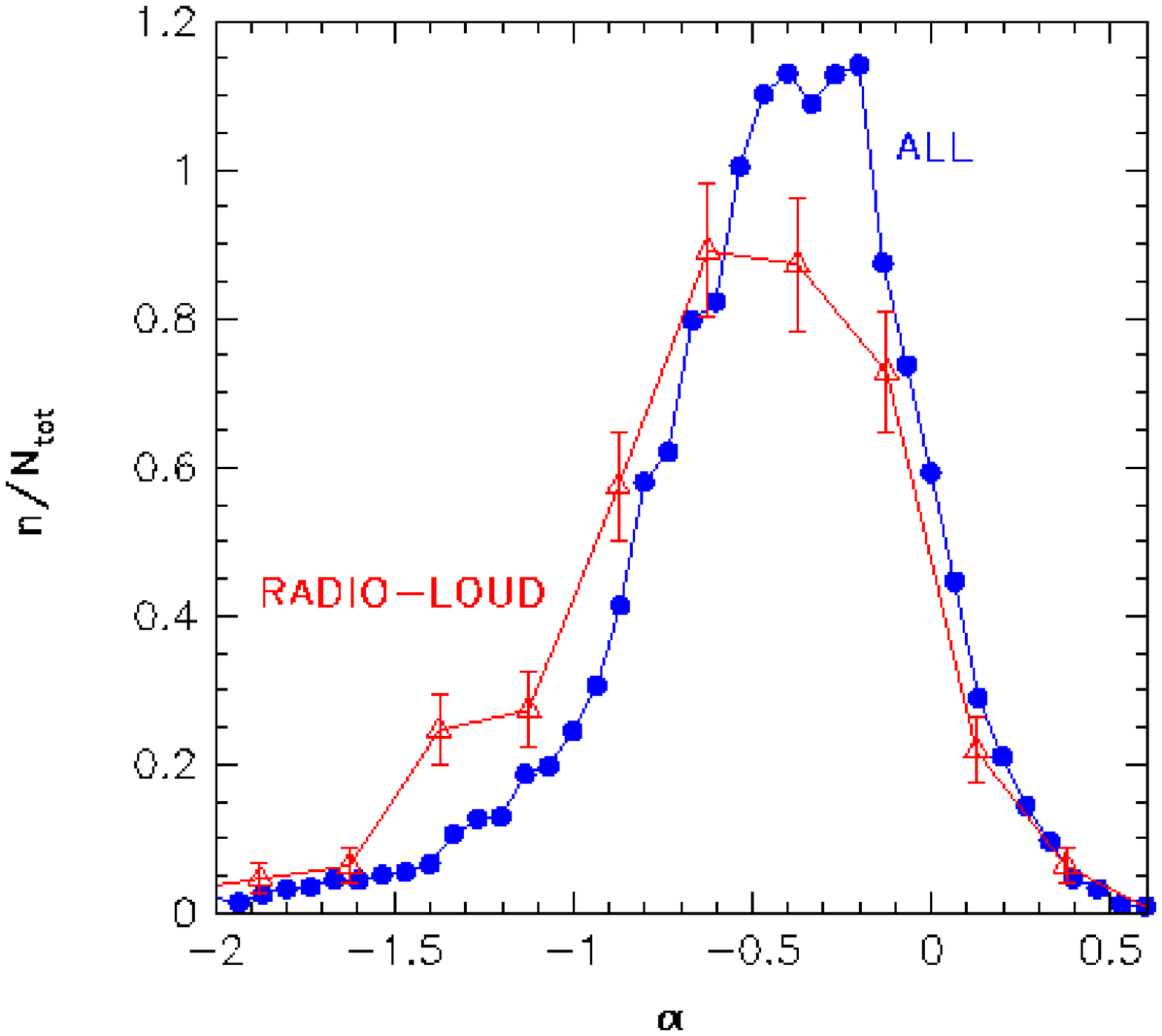,width=8.5cm}
 \caption{Left: the distribution of the quasars continuum power-law index $\alpha$
from the combined fall and spring samples. The dotted line represents the distribution of spectral indices from a sample of quasars at redshift between 2 and 3.5 from Francis (1996). Right: the distribution of spectral indices derived by Ivezi\'c et al. (2002) for about 6800  SDSS quasars
with magnitude brighter than i$=$19. The $\alpha$ distribution for a subsample of 440 radio-loud quasars, indicated by the triangles, is skewed towards more negative values.}
\end{figure*}

In Figure 4 we present the distribution of $\alpha$
from the combined Fall and Spring sample.
The distribution is essentially confined between 0.2 and -1.1 (with one object at -1.8), with
an average spectral index  of $\alpha=-0.57$ and 1$\sigma$ dispersion of $0.33$.
The mean was
weighted by the errors on the individual $\alpha$ measurements.
The median value is $\alpha=-0.51$. The distribution
is not perfectly symmetric, but slightly skewed towards steeper indices.
In Figure 4 (right panel) we also show for comparison the distribution of spectral indices derived by Ivezi\'c \etal\ (2002) for about 6800 SDSS quasars with magnitude brighter than i=19, spanning a large range of redshifts (see also discussion in Richards \etal\ 2002).
The two distributions appear similar.
The values found for our sample are
consistent with the average power law index
derived from the composite SDSS quasar spectra, spanning a redshift range
0.04$< z < $  4.79 (Vanden Berk \etal\ 2001) which is $\alpha=-0.47$.
This index is valid for a large spectral range, 1216\AA\ $\leq \lambda \leq$ 5000 \AA\
restframe: at longer wavelengths, the slope changes to $\alpha=$-1.58.
A similar result was obtained by Francis (1996) who derived $\alpha$
using photometric estimates, including near-IR photometry,
for a sample of quasars spanning a redshift range 0.33 to 3.67
and with a median redshift of 2.
In Figure 4 we plot the spectral indices derived
by Francis (1996) for all the objects with redshift above 2: for these objects
the average spectral index is $\alpha=-0.49$ (median -0.43).
Natali \etal\ (1998) find a similar result for a sample of bright quasars with redshift up to 2.5, with $\alpha$=-0.65.
However this slope is based on the continuum
shortward of the 3000 \AA\ bump, with a much flatter index at 4000
\AA. One discrepant result comes from Kuhn \etal\ (2001), who derived
the slopes of a sample of bright  quasars
with redshift around 3, using both photometry and spectroscopy. They
find slightly flatter average spectra, with $\alpha= -0.34$
(median -0.29); however we note that their spectral slopes for the
z$\sim$0 comparison sample of quasars are also flatter than what was found
in other studies.

In Figure 5 we plot the derived values for the Fall sample, and compare
them to the spectral indices derived for the same quasars in Fan \etal\ (2001a) from
the low resolution spectra and optical photometry.
The error bars plotted in the figure indicate the 1$\sigma$ error.
Basically all spectral indices
are consistent with those derived from optical data
within 3$\sigma$.
However the indices  derived from the combined near-IR and optical broad band
photometry  tend to be flatter (bluer) than those derived from
the optical data alone and the dispersion of $\alpha$ values
 around the mean is also
somewhat smaller.
The average spectral index derived from the
photometry for the 29 objects of the fall
sample alone is $\alpha=-0.59\pm0.34$ with a median value of $-0.47$,
whereas for the compete fall sample (all 39 quasars) Fan \etal\ (2001a) derived an
average value of $\alpha=-0.88 \pm 0.36$ (note however that they applied weights that
included the error measurement as well as the detection probability function).
Similarly Schneider \etal\ (2001) derived an average
$\alpha = -0.93  \pm 0.31$
from spectra of high redshift SDSS quasars, some of which are included in
our sample.
\\
The apparent discrepacy probably arises because the indices are
derived from different wavelength regions  of the continuum
emission: the indices derived by Fan \etal\ 2001a and by Schneider
\etal (2001) are from the continuum immediately blueward of the
\lya\ up to 9000 \AA\ (or less) observed emission, and thus
include only a few hundred \AA\ in the restframe (1250 to
1600-1800 \AA\ depending on the redshift). Furthermore, part of
the Fe complex in the 1500 - 2000 \AA\ region could make the i and
z band brighter, thus producing a steepening of the spectral
indices. On the other hand our photometry spans a much larger
wavelength  region and, most important, includes longer
wavelengths: at redshift 3.65 (our lower redshift) the photometry
cover from 1500 to  4700 \AA\ (restframe) whereas at redshift 5,
our most distant object, the coverage goes from $\sim$ 1150 \AA\
up to $\sim$ 3700 \AA. Fig 3 illustrates that the continuum
immediately longward of Ly$\alpha$ is somewhat redder than at
longer wavelengths.

We conclude that the continuum properties of the
high redshift quasars  in our sample
at optical restframe wavelengths are comparable to those of their lower
redshift counterparts, with no significant change with epoch.
We also find that the spectral slopes change somewhat
depending on the wavelength regions used to measure them, as already
indicated by the results of Vanden Berk \etal\ (2001) and Natali \etal\ (1998).

\begin{figure}
\psfig{figure=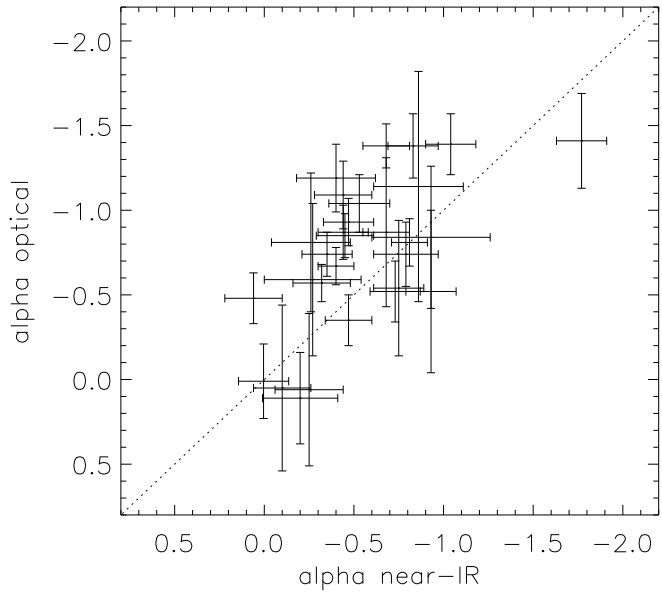,width=11cm}
 \caption{The spectral index of the continuum emission: on the y axis is the
value derived by Fan \etal\ 2001 from the low resolution spectra, on the x-axis
the value derived from optical and near-IR  photometry. The error bars represent the 1 $\sigma$ uncertainties.}

\end{figure}
\section{Summary}
We have presented near-IR photomety (J,H,K$'$) of a sample of 45
color-selected SDSS quasars at 3.6$<$ z $<$5. We have determined
the slopes of the optical continuum at restframe wavelengths 1200
\AA $< \lambda_{rest} <$ 5000 \AA, and we find an average spectral
slope $\alpha = -0.57$ and a 1 $\sigma$ dispersion of $0.33$. This
value is consistent with those found by many authors for lower
redshft quasars at similar restframe wavelengths, so we conclude
that there is no evidence for a redshift evolution of the quasar
continuum properties, as it was previously suggested. By comparing
our slopes with those determined by Fan \etal\ (2001) for the same
objects, using the optical spectroscopy and photometry alone, we
can see that the addition of near-IR data yields a somewhat bluer
continuum and more robust evaluation. This implyes that the
 determination of the spectral slope changes somewhat
depending on the wavelength used to measure them,
as already indicated by previous results.

\begin{acknowledgements}
This paper is based on observations in the framework of the ``Calar Alto Key Project for SDSS Follow-up
        Observations'' (Grebel 2001) obtained at the German-Spanish Astronomical Centre, Calar
        Alto Observatory, operated by the Max Planck Institute for Astronomy, Heidelberg jointly
        with the Spanish National Commission for Astronomy.
\\
Funding for the creation and distribution of the SDSS Archive has been provided by the Alfred P. Sloan Foundation, the Participating Institutions, the National Aeronautics and Space Administration, the National Science Foundation, the U.S. Department of Energy, the Japanese Monbukagakusho, and the Max Planck Society. The SDSS Web site is http://www.sdss.org/.
The SDSS is managed by the Astrophysical Research Consortium (ARC) for the Participating Institutions. The Participating Institutions are The University of Chicago, Fermilab, the Institute for Advanced Study, the Japan Participation Group, The Johns Hopkins University, Los Alamos National Laboratory, the Max-Planck-Institute for Astronomy (MPIA), the Max-Planck-Institute for Astrophys
ics (MPA), New Mexico State University, University of Pittsburgh, Princeton University, the United States Naval Observatory, and the University of Washington.
MAS acknowledges the support of  NSF grant AST 00-71091

\end{acknowledgements}

{}

\begin{thebibliography}{}
\bibitem[Anderson et al.(2001)]{and01} Anderson, S.~F.~et
al.\ 2001, AJ, 122, 503
\bibitem[Barkhouse \& Hall(2001)]{2001AJ....121.2843B} Barkhouse, W.~A.~\&
Hall, P.~B.\ 2001, AJ, 121, 2843
\bibitem[Bechtold et al.(1994)]{1994AJ....108..374B} Bechtold, J.~et al.\
1994, \aj, 108, 374
\bibitem[Elias, Frogel, Matthews, \& Neugebauer(1982)]{1982AJ.....87.1029E}
Elias, J.~H., Frogel, J.~A., Matthews, K., \& Neugebauer, G.\ 1982, \aj,
87, 1029
\bibitem[Fan et al.(2003)]{fan03} Fan, X.~et al.\ 2003, AJ, in press
(astro-ph/0301135)
\bibitem[Fan et al.(2001)]{fan01} Fan, X.~et al.\ 2001a, AJ, 121, 31
\bibitem[Fan et al.(2001)]{2001AJ....122.2833F} Fan, X.~et al.\ 2001b, \aj,
121, 54
\bibitem[Fan et al.(2000)]{fan00} Fan, X.~et al.\ 2000, AJ, 119, 1
\bibitem[Fan et al.(1999)]{fan99} Fan, X.~et al.\ 1999, AJ, 118, 1
\bibitem[Finlator et al.(2000)]{2000AJ....120.2615F} Finlator, K.~et al.\
2000, \aj, 120, 2615
\bibitem[Francis(1996)]{fra96} Francis, P.~J.\ 1996, Publications of the Astronomical
Society of Australia, 13, 212
\bibitem[Fukugita {\em et al.} 1996]{F96}Fukugita, M.~et al.\ 1996, AJ, 111, 1748¡
\bibitem[Grebel 2001]{gre01} Grebel, E. K. 2001, Reviews in Modern Astronomy 14, 223
\bibitem[Gunn et al.~1998]{Gunnetal} Gunn, J.E., et al.~1998, AJ, 116, 3040H
\bibitem[Hall et al.(2002)]{2002ApJS..141..267H} Hall, P.~B.~et al.\ 2002,
ApJS, 141, 267
\bibitem[Hogg et al. 2001]{Hogg01} Hogg, D., et al.~2001, AJ, 122, 2129
\bibitem[Hunt et al.(1998)]{hun98} Hunt, L.~K.~et al.\ 1998, AJ, 115, 2594
\bibitem[Ivezic et al.(2002)]{hun98} Ivezic, Z. ~et al.\ 2002, AJ, 124, 2364
\bibitem[Kuhn, Elvis, Bechtold, \& Elston(2001)]{2001ApJS..136..225K} Kuhn,
O., Elvis, M., Bechtold, J., \& Elston, R.\ 2001, \apjs, 136, 225
\bibitem[Kennefick et al.(1995)]{1995AJ....110...78K} Kennefick, J.~D. et al.\ 1995, \aj, 110, 78
\bibitem[Leggett et al. (2002)]{2002ApJ...564..452L} Leggett, S.~K.~et al.\
2002, ApJ, 564, 452
\bibitem[Natali, Giallongo, Cristiani, \& La
Franca(1998)]{nat98} Natali, F., Giallongo, E., Cristiani,
S., \& La Franca, F.\ 1998, AJ, 115, 397
\bibitem[Pier et al.~2002]{Astrom} Pier, J. et al.~2002, AJ, in press (astro-ph/0211375)
\bibitem[Richards et al. 2002]{Richards02} Richards, G.~T., \etal\ 2002, AJ,  123, 2945
\bibitem[Richards et al.(2001)]{rich01} Richards, G.~T.~et al.\ 2001, AJ, 121, 2308
\bibitem[Richstone \& Schmidt(1980)]{1980ApJ...235..361R} Richstone,
D.~O.~\& Schmidt, M.\ 1980, \apj, 235, 361
\bibitem[Rodriguez Espinosa, Stanga, \aap Moorwood(1988)]{rod88} Rodriguez Espinosa, J.~M., Stanga,
R.~M., \& Moorwood, A.~F.~M.\ 1988, A\&A, 192, 13
\bibitem[Schlegel, Finkbeiner, \& Davis(1998)]{1998ApJ...500..525S}
Schlegel, D.~J., Finkbeiner, D.~P., \& Davis, M.\ 1998, \apj, 500, 525
\bibitem[Schneider et al.(2001)]{sch01} Schneider, D.~P.~et
al.\ 2001, AJ, 121, 1232
\bibitem[Smith et al.(2002)]{smi02}Smith, J.A., \etal\ 2002, AJ, 123, 2121
\bibitem[Stoughton et al.(2002)]{2002AJ....123..485S} Stoughton, C.~et al.\
2002, AJ, 123, 485
\bibitem[Vanden Berk et al.(2001)]{van01} Vanden Berk, D.~E.~et al.\ 2001, AJ, 122, 549
\bibitem[Wainscoat and Cowie]{wai92} Wainscoat R., Cowie L., 1992, AJ 103, 332
\bibitem[Warren, Hewett, \& Osmer(1994)]{1994ApJ...421..412W} Warren,
S.~J., Hewett, P.~C., \& Osmer, P.~S.\ 1994, \apj, 421, 412
\bibitem[Wills, Netzer, \& Wills(1985)]{1985ApJ...288...94W} Wills, B.~J.,
Netzer, H., \& Wills, D.\ 1985, \apj, 288, 94
\bibitem[York et al.(2000)]{york00} York, D.~G.~et al.\ 2000,
AJ, 120, 1579
\bibitem[Zheng et al.(2000)]{zhe00} Zheng, W.~et al.\ 2000, AJ, 120, 1607


\end{thebibliography}
\end{document}